# Topological characterization of antireflective and hydrophobic rough surfaces: are random process theory and fractal modeling applicable?

Claudia Borri, Marco Paggi[*]

*IMT Institute for Advanced Studies Lucca, Research unit MUSAM – Multi-scale Analysis of Materials, Piazza San Francesco 19, 55100 Lucca, Italy*

email: claudia.borri@imtlucca.it, marco.paggi@imtlucca.it

**Abstract**

The random process theory (RPT) has been widely applied to predict the joint probability distribution functions (PDFs) of asperity heights and curvatures of rough surfaces. A check of the predictions of RPT against the actual statistics of numerically generated random fractal surfaces and of real rough surfaces has been only partially undertaken. The present experimental and numerical study provides a deep critical comparison on this matter, providing some insight into the capabilities and limitations in applying RPT and fractal modeling to antireflective and hydrophobic rough surfaces, two important types of textured surfaces. A multi-resolution experimental campaign by using a confocal profilometer with different lenses is carried out and a comprehensive software for the statistical description of rough surfaces is developed. It is found that the topology of the analyzed textured surfaces cannot be fully described according to RPT and fractal modeling. The following complexities emerge: (*i*) the presence of cut-offs or bi-fractality in the power-law power-spectral density (PSD) functions; (*ii*) a more pronounced shift of the PSD by changing resolution as compared to what expected from fractal modeling; (*iii*) inaccuracy of the RPT in describing the joint PDFs of asperity heights and curvatures of textured surfaces; (*iv*) lack of resolution-invariance of joint PDFs of textured surfaces in case of special surface treatments, not accounted by fractal modeling.

---

[*] Corresponding author (M. Paggi). E-mail: marco.paggi@imtlucca.it; tel: +39 0583 4326 604
Fax: +39 0583 4326 565

**Nomenclature and physical dimensions of the various parameters**

| | |
|---|---|
| $\delta$ | Sampling interval (μm) |
| $L$ | Specimen lateral dimension (μm) |
| $x, y$ | In plane coordinates of a generic sampled height (μm) |
| $z$ | Generic surface height or elevation (μm) |
| $\omega = 2\pi/\delta$ | Sampling frequency (1/μm) |
| $\Phi$ | Profile power spectral density function (PSD) (μm$^3$) |
| $G$ | PSD roughness parameter (μm) |
| $D_p$ | Profile fractal dimension (–) |
| $D$ | Surface fractal dimension (–) |
| $m_0$ | First moment of the PSD, variance of the surface heights (μm$^2$) |
| $m_2$ | Second moment of the PSD, variance of the profile slopes (–) |
| $m_4$ | Fourth moment of the PSD, variance of the summit curvatures (–) |
| $\alpha = \dfrac{m_0 m_4}{m_2^2}$ | Bandwidth parameter (–) |
| $\kappa_1, \kappa_2$ | Summit principal curvatures (1/μm) |
| $\kappa = \sqrt{\kappa_1 \kappa_2}$ | Geometric mean curvature (1/μm) |
| $g = \kappa/\sqrt{m_4}$ | Dimensionless geometric mean curvature (–) |
| $s = z/\sqrt{m_0}$ | Dimensionless asperity height (–) |
| $P(s,g)$ | PDF of summit curvatures and heights according to RPT |
| $P(g)$ | PDF of all asperity curvatures according to RPT |
| $P_{\text{beta}}(g)$ | Proposed PDF of all asperity curvatures |
| $B$ | Beta function |
| $\mu, \gamma, \beta$ | Parameters of the PDFs |

## 1. Introduction

The description of roughness of natural and artificial surfaces has received a great attention by engineers and physicists due to its relevance in tribology. Rough surfaces show property of randomness, i.e., they appear disordered and irregular. Historically, surface textures were commonly measured by contact profilometers by moving a stylus over a length $L$ over the surface. By this analysis, the continuous function of heights, $z(x)$, is sampled at discrete intervals of length, $\delta$, obtaining a discrete curve $z(x_i)$, where $i=1, \ldots, N=L/\delta$, being $N$ the total number of sampled heights. The theory of random process (RPT in the sequel) was applied to such data by extending well-established mathematical methods previously applied to random noise signals [1]. The extension from a 1D signal to 2D surface data was later proposed by Longuet-Higgins [2]. Further developments on the necessity of distinguishing between asperities (maxima of the 3D surface) and peaks (maxima in 2D profiles) for contact problems were proposed by Nayak [3]. Relying on the RPT, distribution functions of asperity heights and curvatures and their joint probability have been derived in [4-6]. Meanwhile, theories for the solution of the normal contact problem between nominally rough surfaces emerged by exploiting the idea of incorporating the statistical description of roughness with the constitutive Hertzian equations to model contact at the asperity level (see, e.g., [7-9], among others).

After the pioneering work by Majumdar and Bhushan [10], the property of statistical self-affinity of surface roughness, i.e., roughly speaking, one part of the surface can be transformed into another by reproducing itself almost exactly under a suitable scaling, emerged and fostered new research and debates, especially on the resolution dependency of statistical parameters and contact predictions [10-13]. Fractal geometry opened also the use of surface simulation methods to numerically generate synthetic rough surfaces to be passed as input of contact theories. Notable examples are the Weierstrass-Mandelbrot function [14-16], the spectral synthesis method [15, 16] and the random midpoint displacement method [17, 18, 20]. Often tacitly considered as representative of real surfaces, these generated height fields have been used to carry out extensive comparisons between microscopical contact theories *à la* Greenwood and Williamson and other numerical methods like FEM or BEM for the solution of the normal contact problem, in order to assess their reliability [21]. So far, two important issues have not yet been fully investigated in the literature and motivate the present study. The first open problem regards the checking of the capabilities and accuracy of statistical models based on the RPT to describe the probability distribution functions (PDFs) of asperity heights and curvatures of numerically generated random fractal surfaces. The second issue regards the applicability of such theories based on random process to real surfaces. Although the agreement with experimental data was fairly good in case of metallic surfaces subjected to different types of machining (spark-machined, grit-blasted, sand-blasted surfaces, see, e.g., [6]), the applicability to textured surfaces should be a matter of discussion. This research is motivated by the fact that in the recent years we have seen a rise in designing and tailoring surface roughness in order to optimize adhesion in biological systems [22-24], capillary forces promoting the movement of liquids [25, 26], thermal and electric contact conductance [18, 19], frictional effects [20, 27], wave transmission [28] and also light reflectance properties fundamental for high efficiency solar energy conversion [29]. In most of such cases, surfaces present complex textures over multiple scales, probably far beyond the capabilities of models based on the RPT and also not just simply fractals.

To provide an insight into these open problems, a comprehensive software for the statistical and fractal characterization of rough surfaces has been developed. The features of this analysis software are illustrated in section 2, together with the multi-resolution experimental methods based on the use of a non contact confocal profilometer. To keep the aims of the present article within reasonable limits, we restrict our attention to the recent elliptical model of rough surface contact by Greenwood [9] among the various statistical models based on the RPT available in the literature. The most relevant results of the critical comparisons are presented in sections 3 and 4. Section 3 focuses on the check of the applicability of RPT to numerically generated fractal surfaces with different fractal dimension and resolution. Section 4.1 focuses on the characterization of textured surfaces used for renewable energy harvesting, that is, polycrystalline silicon with a special type of antireflective coating employed in solar cells. Natural hydrophobic surfaces of Gingko Biloba are analyzed in section 4.2, as representative of surfaces displaying the so-called Lotus effect. Finally, a critical discussion on the obtained results and further open lines of research emerging from the conclusions are presented in section 5.

## 2. Experimental methods and main features of the developed surface analysis code

In order to provide a comprehensive description of the topology of rough surfaces from the statistical and the fractal point of views, a software written in Matlab has been developed. The input has to be specified as a simple text file of three columns containing the $x$, $y$ and $z$ coordinates of the surface height field.

The height field can be obtained from computer-generated fractal surfaces, as those realized by using the random midpoint displacement algorithm [30]. Alternatively, a real surface height field obtained from confocal/interferometric profilometers or from atomic force microscopes can be analyzed. In the present study we use the Leica DCM-3D contactless confocal profilometer available in the laboratory of the Research Unit MUSAM −Multi-scale Analysis of Materials− of the IMT Institute for Advanced Studies Lucca equipped with different lenses providing various sample magnification (10 x, 20 x, 100 x) for multi-resolution analysis. The spatial resolution in the $x$, $y$ and $z$ directions depends on the magnification. For the 10 x magnification, the lateral (along $x$ and $y$) resolution is 1.66 μm and the vertical resolution (along $z$) is 2 μm. For the 20 x magnification the lateral resolution is 0.83 μm and the vertical one is 1 μm. For the 100 x magnification, the lateral resolution is 0.166 μm and the vertical one is 0.2 μm.

The analysis code can perform both 2D and 3D statistical characterizations of the height field. Regarding the 2D analysis, individual profiles can be extracted from the height fields. The statistical distribution of the profile heights is estimated from the histogram data and all the related statistical parameters are determined (mean value, root mean square value, skewness and kurtosis). Moreover, the peaks are identified as the local maxima of the profile. The profile slopes and the peak curvatures are computed via standard finite difference schemes (central or forward finite difference schemes for the slopes and three-point finite difference scheme for the curvatures) and the corresponding statistical distributions and their parameters (mean value, root mean square value, skewness and kurtosis) are computed. Regarding the 3D analysis, asperities whose shape is assumed to be mildly elliptical are identified according to the method proposed by Greenwood [9], which consists in the preliminary computation of the principal curvatures $\kappa_1$, $\kappa_2$ at any surface height:

$$\kappa_1 = \begin{cases} -2\dfrac{-z_{i-1,j}+2z_{i,j}-z_{i+1,j}}{-x_{i-1,j}^2+2x_{i,j}^2-x_{i+1,j}^2}, & \text{if } z_{i,j}>z_{i-1,j} \text{ and } z_{i,j}>z_{i+1,j} \\ 0 & \text{otherwise} \end{cases}$$

$$\kappa_2 = \begin{cases} -2\dfrac{-z_{i,j-1}+2z_{i,j}-z_{i,j+1}}{-y_{i,j-1}^2+2y_{i,j}^2-y_{i,j+1}^2}, & \text{if } z_{i,j}>z_{i,j-1} \text{ and } z_{i,j}>z_{i,j+1} \\ 0 & \text{otherwise} \end{cases} \quad (1)$$

Afterwards, the geometric mean curvature $\kappa=\sqrt{\kappa_1\kappa_2}$ is determined, which has the advantage to help distinguishing between maxima and saddle points of the surface. Using equation (1), all the cases where $\kappa$ is different from zero identify asperities.

The code also provides the first three moments of the spectral density function, i.e. the variance of the asperity heights, $m_0$, the variance of the profile slopes, $m_2$, and the variance of the asperity curvatures, $m_4$. Here, the parameters $m_0$, $m_2$ and $m_4$ refer to statistical properties determined from the empirical distribution functions computed for a given finite (non zero) sampling interval δ. The values are used to compute the dimensionless asperity heights $s=z/\sqrt{m_0}$ and the dimensionless asperity geometric mean curvatures $g=\kappa/\sqrt{m_4}$. From the input height field, the proposed analysis code allows the computation of the joint PDF $P(s,g)$ of dimensionless asperity heights and curvatures. This feature is particularly important, since it allows, for the very first time, the comparison with models based on the random process theory. In particularly, Greenwood [9] has recently proposed a simplified version of the joint PDF $P(s,g)$ in case of mildly elliptical asperities, which is a condition often observed in practice. His formula that will be compared with our experimental results has the following form:

$$P(s,g)=\frac{9}{2\sqrt{2\pi}}\sqrt{\frac{\alpha}{\alpha-1}}g^3\,\mathrm{erfc}\left[\mu\left(3g-\frac{s\alpha^{1/2}}{\alpha-1}\right)\right]\exp\left[\frac{1}{2}\left(3g^2-\frac{\alpha s^2}{\alpha-1}\right)\right] \quad (2)$$

where $\alpha=\dfrac{m_0 m_4}{m_2^2}$ is the bandwidth parameter and $\mu=\sqrt{\dfrac{(\alpha-1)}{(4\alpha-6)}}$. The PDF of all asperity curvatures obtained by integrating equation (2) over all the asperity heights leads:

$$P(g)=\frac{9}{2}g^3\exp\left(\frac{3}{2}g^2\right)\mathrm{erfc}\left(\frac{3}{2}g\right) \quad (3)$$

In addition to the statistical characterization, the power spectral density function (PSD) of the surface profiles is evaluated according to a fast Fourier transform of the related height fields, see [30]. We consider here the computation of the PSD just in two orthogonal directions, averaging over all the surface profiles. The orientation is chosen either generic to prove anisotropy by comparing the corresponding PSDs, or *ad hoc* along specific directions of anisotropy visually identified by optical or scanning electron microscopy. In case of a fractal surface, the profile PSDs are expected to have a power-law dependency on the frequency ω [16]:

$$\Phi = G^{2D_p - 2} \varpi^{-(5 - 2D_p)}, \text{ with } 1 < D_p < 2 \tag{4}$$

From the PSD function's linear fitting in a bi-logarithmic plane, the code evaluates the parameters $D_p$ and $G$. In case of isotropic surfaces, the surface fractal dimension is computed as $D = D_p + 1$.

## 3. Numerically generated fractal surfaces

In this section we check the RPT against the statistical distributions of random fractal surfaces numerically generated according to the random midpoint displacement (RMD) algorithm by considering different surface fractal dimensions and resolutions. The standard RMD method, described in [30], allows generating isotropic surfaces with a PSD of power-law type without any cut-off. The empirical PDFs obtained from the statistical analysis of the generated height fields are compared with the model predictions by equation (2).

*3.1 The effect of the surface fractal dimension*

Square surfaces with lateral size of 1 μm and 257 heights in both the *x* and *y* directions are numerically generated by varying the fractal dimension *D*. The PSD curves of the profiles extracted from two orthogonal directions are almost overlapping each other, as expected for isotropic surfaces. In such a case, it is possible to relate the surface fractal dimension to the profile fractal dimension via $D = D_p + 1$. The PSD curves are plotted vs. ω in figure 1 and the *a posteriori* computed fractal dimension *D* reported in table 1 (measured value, 7th column) is close to the values used as input for the generation algorithm (expected value, 1st column), demonstrating that the RMD method provides a satisfactory approximation of the desired fractal topology. As theoretically expected, the increase of the fractal dimension leads to a much rougher surface with increased variances of asperity heights, slopes, and curvatures.

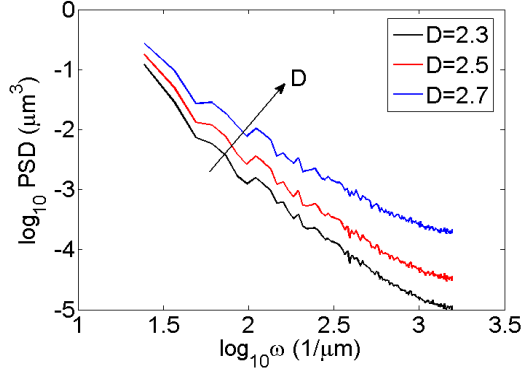

Figure 1: PSD of three numerical generated surfaces with fractal dimension $D$=2.3, 2.5, 2.7.

| D (expected) | $\delta$ | $m_0$ | $m_2$ | $m_4$ | $\alpha$ | D (measured) | G |
|---|---|---|---|---|---|---|---|
| 2.3 | $7.8\times10^{-3}$ | 2.24 | $5.09\times10^2$ | $3.36\times10^7$ | $2.44\times10^2$ | 2.45 | $1.06\times10^1$ |
| 2.5 | $7.8\times10^{-3}$ | 3.63 | $2.98\times10^3$ | $2.51\times10^8$ | $7.02\times10^1$ | 2.59 | 8.62 |
| 2.7 | $7.8\times10^{-3}$ | 6.82 | $2.00\times10^4$ | $2.04\times10^9$ | $1.74\times10^1$ | 2.74 | 4.61 |

Table 1: Statistical and fractal parameters of the numerically generated fractal surfaces with different fractal dimension.

The normalized asperity height and the dimensionless asperity curvature histograms of the surfaces having the smallest and largest values of $D$ are shown in figure 2. The corresponding empirical joint PDF determined from those histograms is displayed by contour lines corresponding to different probability values $P(s,g)$ in a plot which has the same form as that proposed by Greenwood [9]. These results will be referred to as *numerical* in the sequel, to distinguish them from the predictions according to RPT that will be called *theoretical* and that are displayed in an analogous contour plot above the previous one in the same figure. The same color of the numerical and theoretical contour levels corresponds to the same probability value.

It is interesting to note that equation (2) is solely dependent on the bandwidth parameter, which is decreased by one order of magnitude by passing from $D$=2.3 to $D$=2.7, see table 1. Low values of $\alpha$ lead to a slight asymmetry in the theoretically predicted joint PDF, which is a trend also observed in the actual joint PDF of the numerical fractal surfaces. Although the numerical and theoretical contour levels have a shape in fair good agreement with each other, the spacing between the numerical and theoretical contour levels is very different. In general, numerically generated fractal surfaces have a much less steep variation in the probability by changing the dimensionless asperity heights and curvatures as compared to what predicted by RPT for the same bandwidth parameter.

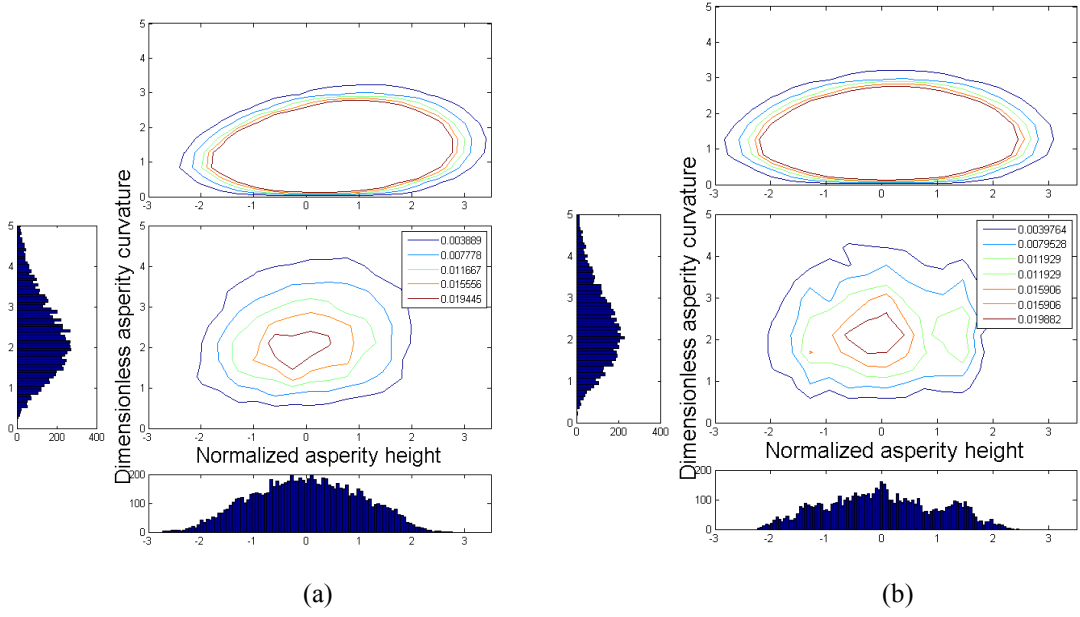

(a)                                          (b)

Figure 2: Comparison between numerical and theoretical (contour plot on the top) joint PDFs of dimensionless asperity heights and curvatures for fractal surfaces with different fractal dimension $D$. (a) $D$=2.3; (b) $D$=2.7.

*3.2 The effect of the surface resolution*

Fractal surfaces with $D$=2.7, lateral size of 1 μm and different resolution (129, 257, 513 or 1025 heights per side) are examined. Since the RMD algorithm is generating square surfaces by successively refining an initial mesh by a successive addition of a series of intermediate heights, surfaces with higher number of heights per side contain their coarser realizations. Therefore, a multi-resolution analysis of RMD surfaces is usually intended to provide a guideline on how the statistical properties of real fractal surfaces should vary by changing the magnification of the profilometer used for the surface height acquisition. Much less is said in the literature as regards the variation of the PDFs of heights and curvatures by varying the resolution.

The profile PSD curves depicted in figure 3 show that the parameter $G$ and $m_0$ are almost resolution independent, whereas a strong influence of δ on $m_2$ and $m_4$ is noticed since they depend on the high frequency cut-off which increases by passing from 129 to 1025 heights per side, consistently with theoretical considerations [10, 13, 31]. We note a slight deviation from the straight line in the high frequency range, which may affect the computation of $D$, especially for low resolutions. Values not too far from the expected one ($D$=2.7) are obtained in all the cases by restricting the best-fitting to the data points before the departure from the linear trend in the bi-logarithmic diagram. The reason for this deviation has to be ascribed to a numerical shortcoming of the RMD algorithm used to generate the surfaces. Contrary to the spectral synthesis method (SSM), which generates the height field via an inverse Fourier transform of the PSD spectrum which is imposed exactly, the RMD provides only an approximation of the power-law PSD function. The RMD algorithm operates by refining an initial coarse height field by adding intermediate heights in the grid with an average elevation computed from those of the neighboring nodes, plus a random increment determined from a Gaussian distribution with a suitably rescaled variance, see [30] for more details. The advantage of RMD over SSM regards the fact that the height field of a surface with a given resolution is contained in its coarser representation, as it happens in case of sampling the same real surface with different focal diameters. In case of the SSM algorithm, surfaces with the same *D* and different resolution would look completely different. Therefore, we prefer here to use the RMD algorithm for a better comparison with the trends obtained from experimental data.

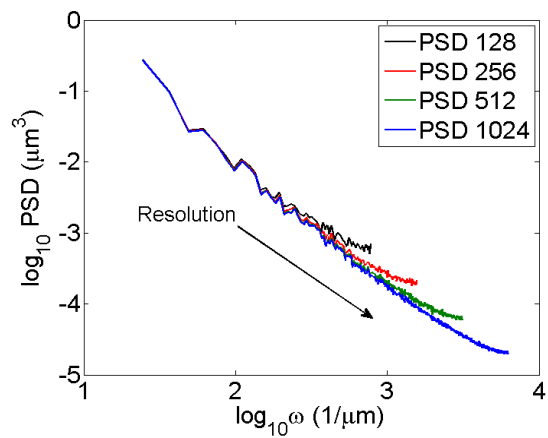

Figure 3: PSD of numerically generated fractal surfaces with surface fractal dimension *D*=2.7 and different spatial resolutions (129, 257, 513 or 1025 heights per unit side).

| $\delta$ | $m_0$ | $m_2$ | $m_4$ | $\alpha$ | D (measured) | G |
|---|---|---|---|---|---|---|
| 1/128 | 6.83 | $7.62 \times 10^3$ | $1.92 \times 10^8$ | $1.12 \times 10^1$ | 2.70 | 7.08 |
| 1/256 | 6.82 | $2.00 \times 10^4$ | $2.04 \times 10^9$ | $1.74 \times 10^1$ | 2.74 | 4.61 |
| 1/512 | 6.81 | $5.30 \times 10^4$ | $2.17 \times 10^{10}$ | $2.64 \times 10^1$ | 2.72 | 5.29 |
| 1/1024 | 6.81 | $1.78 \times 10^5$ | $2.92 \times 10^{11}$ | $2.59 \times 10^1$ | 2.70 | 5.72 |

Table 2: Statistical and fractal parameters of the numerically generated fractal surfaces with different resolution.

From the examination of the numerical and theoretical PDFs corresponding to the coarsest and the finest surfaces, figure 4, we note that the numerical results become smoother and smoother as long as the size of the asperity population increases by refining the surface. However, the probability contour levels have almost the same spacing and location regardless of the resolution. This is quite important if we consider that RMD surfaces with different resolutions have a tremendous different normal contact response, as numerically proven in [11]. More specifically, they found that the finer the surface representation, the smaller the real contact area for a given normal contact pressure. Since the present results pinpoint that the statistical distributions of the dimensionless asperity heights and curvatures are almost unaffected by the resolution, this implies the important result that the different contact response of fractal surfaces can be ascribed solely to the change in the parameters $m_2$ and $m_4$, i.e., by the tail of the PSD for high frequencies.

Again, as in figure 2, from the comparison between the actual joint probability contour plots of the numerically generated surfaces and those of the theoretical ones we note that the distribution of asperity heights of fractal surfaces is more symmetric and presents much narrower tails than what predicted by RPT.

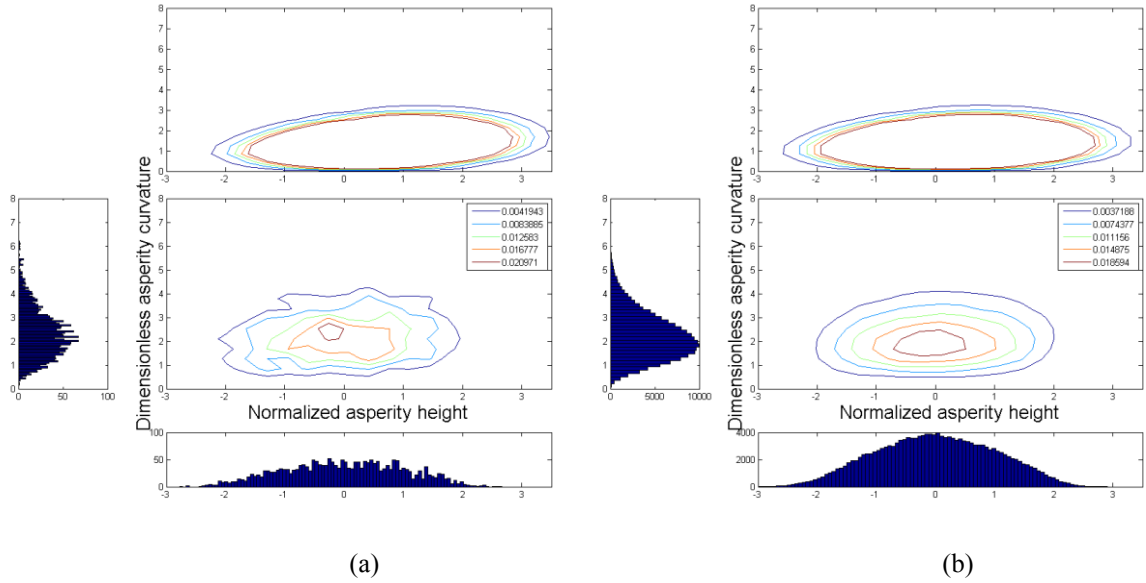

(a)           (b)

Figure 4: Comparison between numerical and theoretical (contour plot on the top) joint PDFs of dimensionless asperity heights and curvatures for fractal surfaces with *D*=2.7 and different resolutions *δ*. (a) *δ*=1/128; (b) *δ*=1/1024.

**4. Analysis of artificially textured surfaces**

*4.1 Silicon solar cells with anti-reflecting coating*

We examine in this section polycrystalline Silicon solar cells with anti-reflecting coating realized by sand-blasting the Silicon wafer to obtain desired roughness on its surface. Subsequently, the wafer is immersed in an acid-etching solution to form a uniform surface texture with low reflectance, see the surface image in Fig. 5 obtained using the scanning electron microscope (SEM) Zeiss EVO MA15 available in the laboratory of the research unit MUSAM at the IMT Institute for Advanced Studies Lucca.

The morphological features of rough surfaces are relevant for their reflectance properties, as pointed out in [32]. The topic is still quite open since most of the theoretical contributions in the literature are assuming either a Gaussian probability distribution function of surface heights [33] or simplified pyramidal models with a single lateral size of the pyramidal asperities. Alternatively, a uniform distribution for the lateral size of the pyramidal asperities with a very limited scale range has been proposed in [34].

During the last few years in the field of photovoltaics, novel etching techniques by varying methodology, time and acid concentration have been proposed to realize pyramidal textures with a much wider range of lateral sizes [35], random patterns [29], or with shallow pits with a worm-like morphology [36] like those analyzed in the present study. The evaluation of the real surface area, which is clearly a multi-scale property dependent on the number of length scales of roughness, influences both reflection [36], which is drastically diminishing by increasing the real surface area in acid etched multi-crystalline silicon, and the applicability of the electrochemical capacitance-voltage method, which also requires the knowledge of the real surface area to profile the surface doping concentration of silicon solar cells [37]. Moreover, a complete multi-scale characterization is useful for the numerical generation of rough surfaces matching the properties of the real ones for subsequent ray tracing simulations.

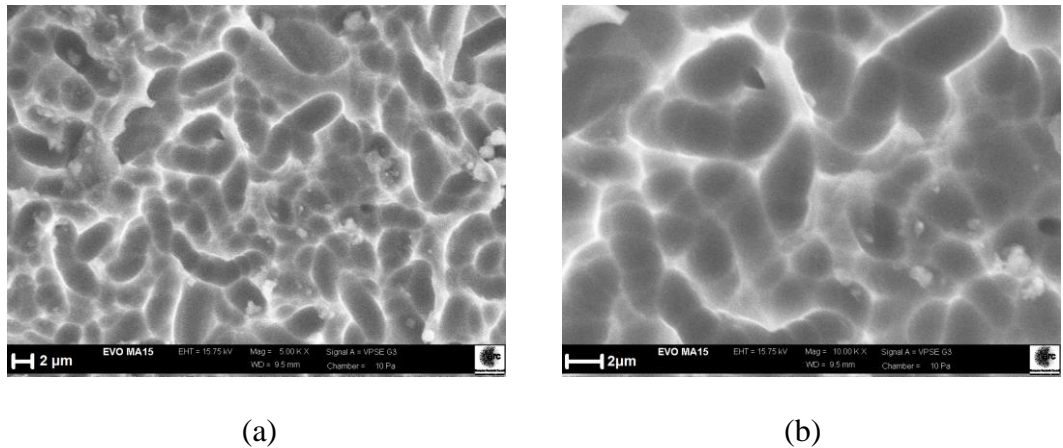

(a)          (b)

Figure 5: SEM images of the textured Silicon solar cell surface at different magnifications. (a): VPSE, 15.75 kV, 5000 x, 9.5 mm working distance; (b): VPSE, 15.75 kV, 10000 x, 9.5 mm working distance.

A multi-resolution analysis of the height field is herein carried out by acquiring the surface with the confocal profilometer at different magnifications (10 x, 20 x, 100 x), see figure 6. Each squared sampled area has 512 heights per side. By increasing the magnification, the textured wavy asperities become clearly visible (see the red areas of higher elevation in figure 6(c)).

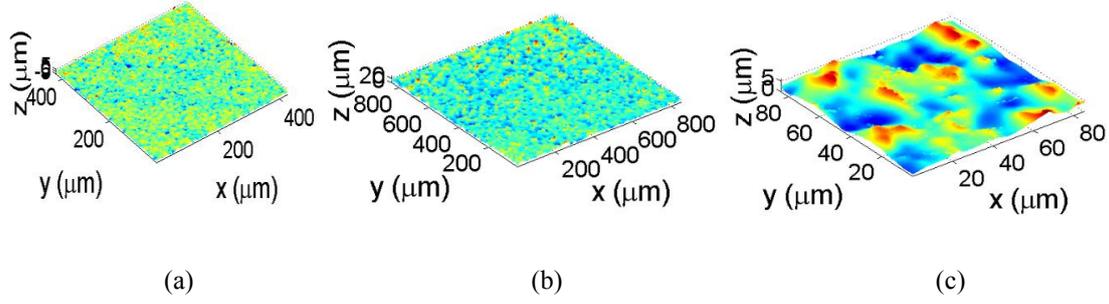

(a)                  (b)                (c)

Figure 6: Surface topography of Silicon solar cells sampled with the confocal profilometer at different magnifications.

Since the surface presents a complex random structure, its roughness characterization is not trivial. As compared to fractal surfaces, cut-offs to the power-law trend of the PSD may take place. Moreover, results may depend on the size of the sample which should be chosen large enough to be statistically representative of the whole surface. To address these two issues, we use different magnifications and we analyze surfaces of different size by collecting various patches together (stitching).

From the analysis of the PSDs, the surface was found to be almost isotropic, so that we show in figure 7 the PSDs for a generic direction. The multi-resolution analysis in figure 7(a) with 10 x and 100 x magnifications clearly shows a power-law trend of the PSD for the high frequency range and the appearance of a cut-off to the power-law regime for $\log_{10}\omega \sim -0.5$ 1/μm, i.e., for a length scale of about 20 μm. Note that this is the typical size of a wavy asperity of the textured surface, as it can be noticed from figure 6(c). Therefore, the fractal regime is almost related to micro-roughness present at scales smaller than the lateral size of the wavy asperity realized by sand-blasting.

By increasing the surface size by stitching $n \times n$ square samples of 512 heights per side each, the PSD curves tend to converge towards a single one, especially for the highest resolution, see figure 7(b). The same convergence can be noticed from the table of the statistical parameters (table 2). The surface fractal dimension of about 2.3 has been estimated by fitting the PSD only within its power-law regime. The analysis of figure 7 also shows a trend not observed in case of fractal modeling, that is, PSDs corresponding to higher resolutions (100 x) are slightly shifted with respect to those with lower resolutions (10 x) towards higher values.

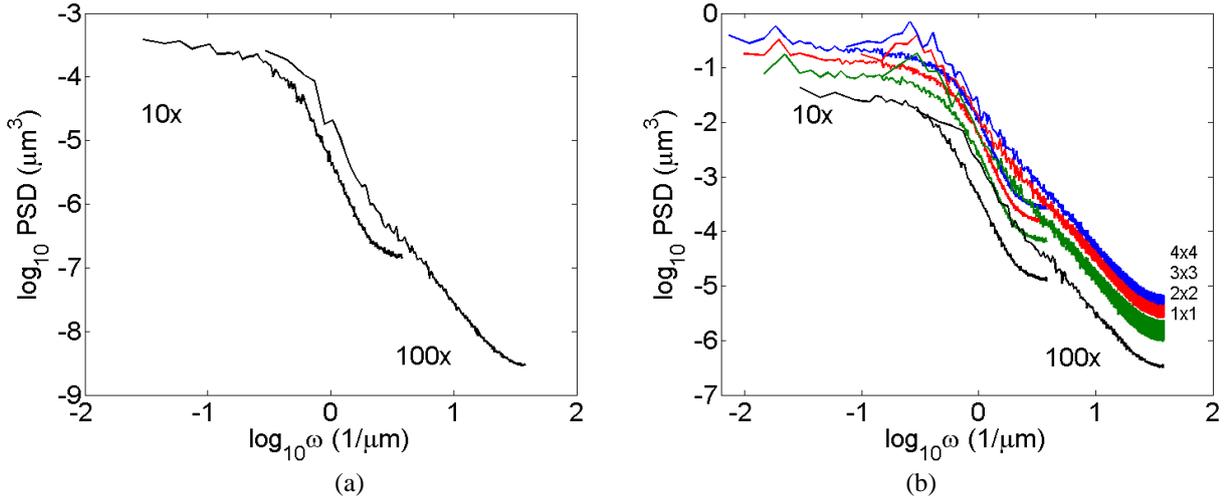

(a) (b)

Figure 7: PSD curves for Silicon textured surfaces with 10 x or 100 x magnifications and different stitching sizes (1 × 1, 2 × 2, 3 × 3 and 4 × 4).

| Size | δ | $m_0$ | $m_2$ | $m_4$ | α | D | G |
|---|---|---|---|---|---|---|---|
| 100 x | | | | | | | |
| 1x1 | 0.166 | 8.99 | $7.49 \times 10^{-2}$ | $9.88 \times 10^{1}$ | $3.46 \times 10^{3}$ | 2.20 | $1.32 \times 10^{-9}$ |
| 2x2 | 0.166 | 6.74 | $1.50 \times 10^{-1}$ | $1.05 \times 10^{2}$ | $4.70 \times 10^{3}$ | 2.33 | $3.99 \times 10^{-5}$ |
| 3x3 | 0.166 | 5.75 | $1.53 \times 10^{-1}$ | $1.05 \times 10^{2}$ | $4.42 \times 10^{3}$ | 2.37 | $3.26 \times 10^{-4}$ |
| 4x4 | 0.166 | 5.47 | $1.47 \times 10^{-1}$ | $1.07 \times 10^{2}$ | $4.85 \times 10^{3}$ | 2.39 | $1.03 \times 10^{-3}$ |
| 10 x | | | | | | | |
| 1x1 | 1.66 | 0.71 | $1.93 \times 10^{-1}$ | $1.90 \times 10^{-1}$ | 7.29 | 2.35 | $4.45 \times 10^{-5}$ |
| 2x2 | 1.66 | 1.24 | $2.13 \times 10^{-1}$ | $4.32 \times 10^{-1}$ | $1.08 \times 10^{1}$ | 2.36 | $3.85 \times 10^{-4}$ |
| 3x3 | 1.66 | 1.15 | $2.00 \times 10^{-1}$ | $4.18 \times 10^{-1}$ | $1.08 \times 10^{1}$ | 2.38 | $1.34 \times 10^{-3}$ |
| 4x4 | 1.66 | 1.07 | $2.09 \times 10^{-1}$ | $4.83 \times 10^{-1}$ | $1.13 \times 10^{1}$ | 2.38 | $2.64 \times 10^{-3}$ |

Table 3: Statistical and fractal parameters of the silicon solar cell textured surface with different resolution (10 x or 100 x) and size.

The comparison between the experimental and theoretical joint PDFs of dimensionless asperity heights and curvatures is proposed in figure 8. The experimental joint PDF is particularly different from the theoretical one for the same bandwidth parameter. A high frequency of asperities with very small curvature is detected. This trend is even amplified by increasing the magnification, figure 8(b), thus not displaying the resolution-invariance noticed in case of ideal fractals (figure 4).

A closer look at the PDF of the asperity curvatures for values close to zero is done in figure 9, where we can readily see the very sharp shape of the PDF that was not possible to accurately display in the contours of figure 8 with an adequate resolution. RPT predictions by equation (3) are shown with a solid black curve in the same figure. A fitting of the asperity curvature distribution with a beta function is attempted and seems to be promising to interpret the experimental trend. The used equation is a beta prime distribution as in the form of:

$$P_{\text{beta}}(g) = \frac{g^{\gamma-1}(1+g)^{-\gamma-\beta}}{B} \qquad (5)$$

where $B$ is the beta function:

$$B = \int_0^1 g^{\gamma-1}(1-g)^{\beta-1}\,dg \qquad (6)$$

and $\gamma$ and $\beta$ are the best fitting parameters evaluated via a nonlinear regression of the experimental data. For the 10 x magnification we found $\gamma = 1.83$ and $\beta = 4.01$, and for the 100 x magnification $\gamma = 1.02$ and $\beta = 9.83$.

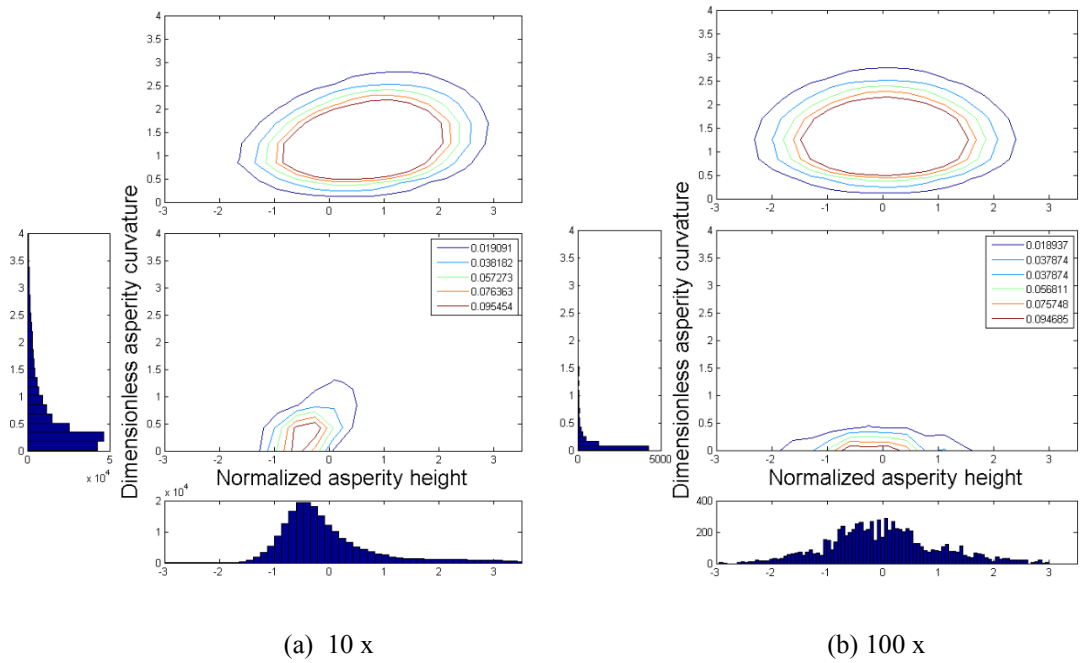

(a) 10 x       (b) 100 x

Figure 8: Comparison between analytical (contour plot on the top) and experimental joint PDFs of asperity heights and curvatures (4 × 4 stitching and two different magnifications).

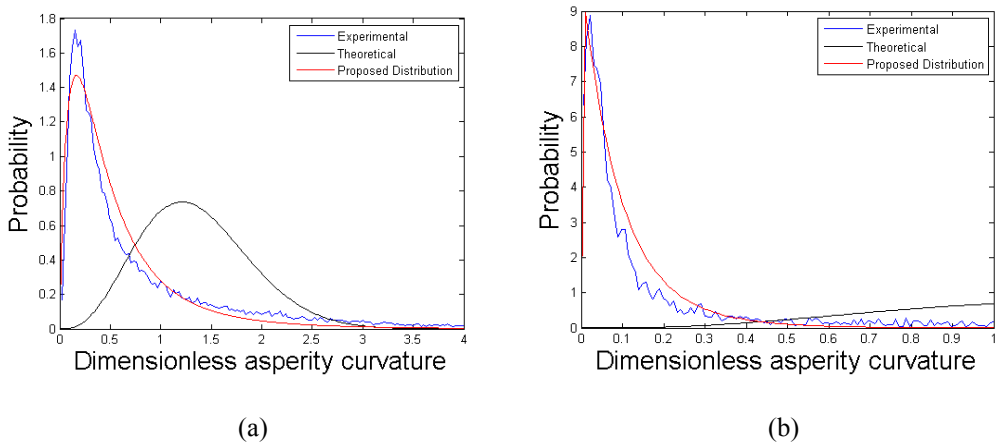

(a)       (b)

Figure 9: PDF of asperity curvatures (4 × 4 stitching and two different magnifications): experimental, theoretical (random process theory) and proposed distribution (beta function).

The reason for these differences from RPT predictions should be ascribed to the special surface treatment consisting in sand-blasting and subsequently acid-etching which leads to many asperities with very small curvatures. Sand-blasting is usually creating very deep roughness and therefore high curvatures. This effect was strongly mitigated by acid-etching which leads to a smoothing out of deep peaks and valleys by removing the blast-damaged layers and creating a uniform texture. Figures 8 and 9 also pinpoint that the effect of acid-etching is more pronounced at higher magnifications than at lower ones, as physically intuitive.

*4.2 Hydrophobic textured surfaces: the Ginkgo Biloba leaf*

In this section we analyze the natural surface of the Ginkgo Biloba leaf. From optical microscope images (figure 10) we observe an alternance of lower (light green) and higher (dark green) rough regions. Both regions are rough and the level of the higher regions is about 25-30 μm above the lower ones. The scope of this texture is mostly to convey water. The hydrophobic property of such leaves (also common to Lotus) is obtained through a microscopical roughness in the form of small asperities of spherical shape that can be observed with the profilometer by increasing the magnification, see figure 11.

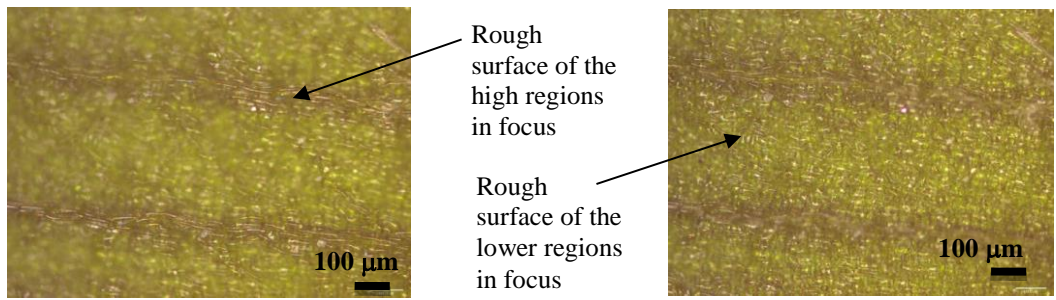

Figure 10: Optical microscope image of Ginkgo Biloba leaf at 10 x magnification. The surface is naturally textured by the alternance of rough lower and higher regions.

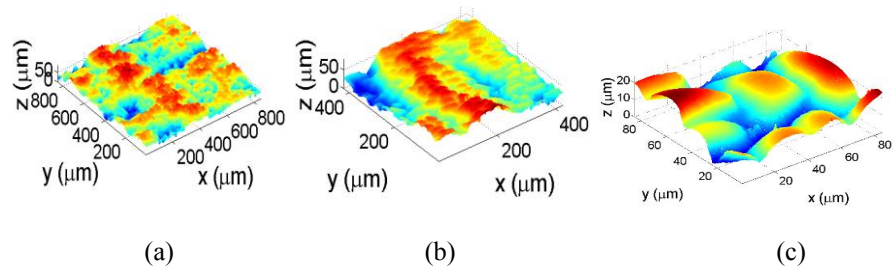

(a)  (b)  (c)

Figure 11: Surface topographies of the Ginkgo Biloba leaf sampled with the confocal profilometer at different

magnifications. At 10 x, high cliffs are visible along the *y* direction. The image at 100 x shows roughness responsible for the Lotus effect on the top of the tallest wavy areas in figure 20 x.

Although it is understood that the hydrophobic properties of surfaces can be modified by surface roughness, natural surfaces like those of Ginkgo Biloba present several length scales and mimicking their topology with artificial texturing is not an easy task. Most of the pioneering attempts have proposed a single scale pillar or groove structure as in [38, 39]. Even an ideal hierarchical configuration with two or three levels of pillars of smaller size mounted on the top of bigger pillars as that modelled in [40] is still a deterministic and regular topology far from the real one. More recently, Kubiak et al. [41] have investigated the wettability of metallic surfaces depending on the abrasion amplitude obtained by polishing. They found that the apparent contact angle was non-monotonically dependent on the variance of surface heights, $m_0$. A complex dependency between contact angles and r.m.s. roughness was also noticed in [42].

Correlations with other statistical parameters dependent on the multi-scale properties of roughness, such as $m_2$ and $m_4$, or with the parameters of the PSD as computed in the present study, could possibly help in distinguishing between different forms of roughness in terms of variation of wettability with polishing.

The PSD in the *y* direction of the surfaces depicted in figure 11 are shown in figure 12. At 10 x and 20 x magnifications, the PSDs are almost overlapping each other. The PSD at 100 x, on the other hand, is shifted towards higher values. This shift can be explained by the particular surface texture consisting of almost elliptical asperities with random orientation leading to a form of waviness, with a very fine superimposed roughness (see figure 11(c)). This morphology was only partially detectable at 10 x and 20 x magnifications. This phenomenon presents some analogies with Silicon wafer surfaces analyzed in [43], where a fine scale waviness was the reason for the shift of the PSD to higher values in the magnification regime proper of AFM.

Although a cut-off to the power-law trend is not observed, the shape of the PSD is not just a single power-law over the whole frequency range, since a change in the slope takes place for $\omega=1$ $\mu m^{-1}$. Therefore, a kind of bi-fractality could be invoked, with $D_1=2.47$ in the low frequency range and $D_2=2.57$ in the high frequency range after kinking.

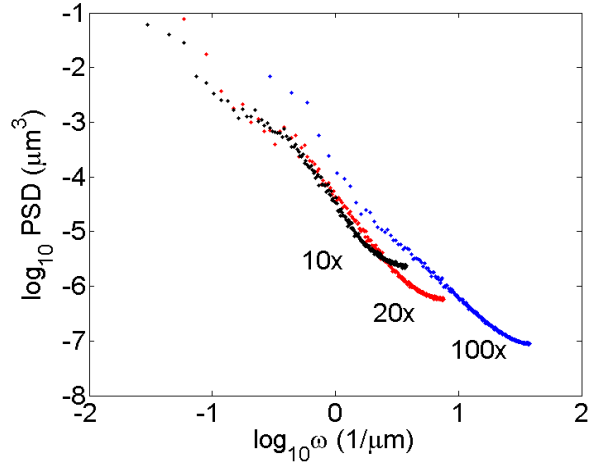

Figure 12: PSD of the Ginko leaf in the *y* direction at different magnifications (10 x, 20 x, 100 x).

A comparison between the experimental and the theoretical joint PDFs of dimensionless asperity heights and curvatures at 10 x and 100 x is proposed in figure 13 and table 3. At low magnifications (10 x, figure 13(a)), the shape of the experimental probability contour levels are conform to what predicted by RPT. A skewness in the asperity height distribution is observed both in theory and in experiments. At a higher magnification (100 x, figure 13(b)), the skewness of the experimental asperity height distribution is reduced, consistently with RPT predictions. However, the distribution of dimensionless asperity curvatures has a shape significantly different from the theoretically expected one, again due to the substantial rise in the frequency of micro-asperities with very small curvatures placed on the top of the lower scale elliptical asperities. Therefore, the joint PDF is not resolution-invariant as it would be expected from fractal modeling (figure 4).

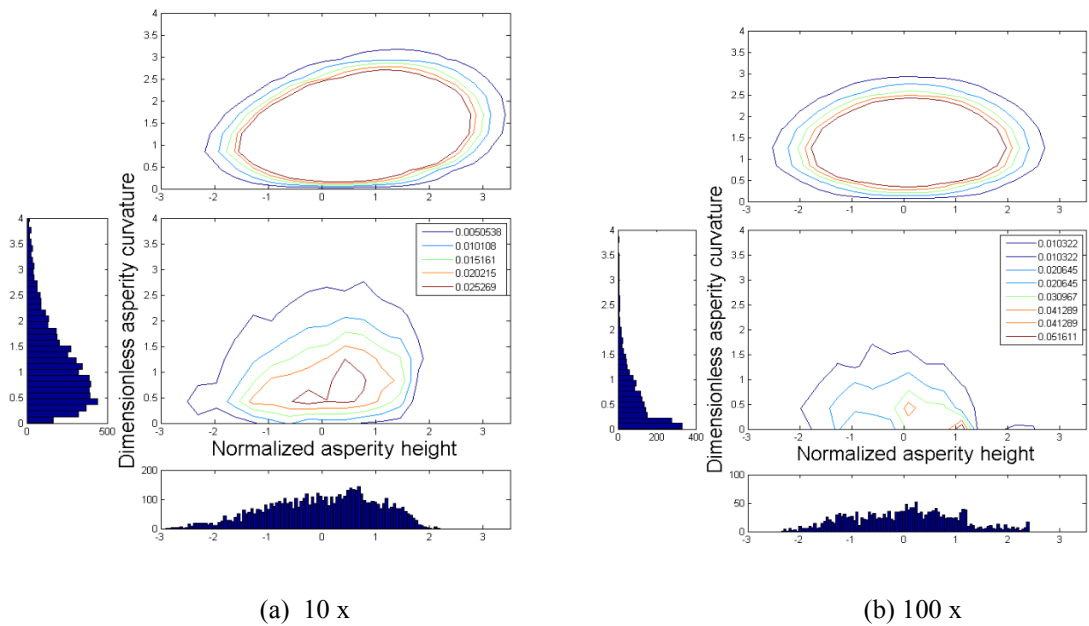

(a) 10 x            (b) 100 x

Figure 13: Comparison between analytical and experimental joint PDFs of asperity heights and curvatures at two different magnifications.

## 5. Discussion and conclusion

In the present study, the joint PDFs for the asperity heights and curvatures of rough surfaces predicted according to the RPT –solely dependent on the bandwidth parameter $\alpha$– have been compared for the very first time with those of numerically generated fractal surfaces and of experimental data obtained from the statistical analysis of the height fields of textured surfaces acquired by using a non contact confocal profilometer. Two types of texturing have been characterized from the statistical point of view, namely an artificially rough coating to increase the anti-reflectance of solar cell surfaces and a natural texturing of Gingko Biloba leaves to enhance the hydrophobic effect via the so-called Lotus effect.

The comparison between RPT predictions and the actual statistics of fractal surfaces shows a general good agreement as far as the joint PDFs for different surface fractal dimensions are concerned. However, although the shape of the contour levels is similar, a closer look shows that the numerically generated fractal surfaces have a much less steep variation in the probability values by varying the dimensionless asperity heights and curvatures, as compared to what predicted by RPT for the same bandwidth parameter. By increasing the surface resolution, the measured probability contour levels become smoother and smoother, but they do not substantially change their shape. This is an important result implying that the well-known change in the slope of the real contact area-load curves observed in [11, 12] by varying the surface resolution has to be ascribed to the change in the spectral moments, rather than to a change also in the dimensionless PDFs of asperity heights and curvatures. Therefore, although statistical parameters are resolution dependent, the use of RPT still holds as far as the description of the shape of the dimensionless PDFs of fractal surfaces is concerned.

On the other hand, the analysis of experimental results concerning textured surfaces shows that both the RPT and fractal modeling are not completely able to describe such complex topologies and for some features they fail.

As regards the surfaces with anti-reflective coating, the PSD presents a cut-off to the power-law regime in the low frequency range in correspondence to a sampling length equal to the characteristic lateral size of the wavy asperities artificially created by sand-blasting. Therefore, roughness with fractal properties can only be observed for scales smaller than such a characteristic size, i.e., where roughness is unaffected by mechanical processing. In addition to the cut-off to the power-law scaling of the PSD not taken into account by fractal modeling, another discrepancy regards the scaling of the PSD by varying the resolution. Fractal modeling suggests almost no shift in the PSD, whereas a significant shift has been noticed in the experimental data. The RPT applied to textured surfaces often fails in predicting the joint PDF of asperity heights and curvatures. In particular, the distribution of dimensionless curvatures is far from being that expected according to RPT, mostly due to the special surface treatment. The large frequency of asperities with very small dimensionless curvatures can be better approximated by a beta distribution, as attempted in the present study.

The multi-resolution analysis of roughness of the Ginkgo Biloba leaves shows that the PSD function cannot be described by a single power-law PSD over the whole frequency range. RPT provides a good representation of the joint PDFs of asperity heights and curvatures at low magnifications (10 x and 20 x), whereas it fails at higher magnifications (100 x) when the statistical distribution of curvatures presents again an unpredicted rise in the frequency of asperities with very small curvatures, thus not confirming the property of resolution invariance of the joint PDF typical of ideal fractals.

The comparisons proposed in the present study are important for the development of new theories improving the description of the statistics of asperity curvatures, which plays an important role in governing the anti-reflective and hydrophobic properties of rough surfaces. Moreover, the results pinpoint the need of developing new algorithms for the numerical simulations of the topology of the complex real textured surfaces that are not just following the scaling of ideal fractals.

**Acknowledgements**

The research leading to these results has received funding from the European Research Council under the European Union's Seventh Framework Programme (FP/2007–2013)/ERC Grant Agreement No. 306622 (ERC Starting Grant ''Multi-field and multi-scale Computational Approach to Design and Durability of PhotoVoltaic Modules'' – CA2PVM).

**References**

[1] Rice SO (1945) Mathematical analysis of random noise, *Bell System Technical Journal*, **24**, 46-156.

[2] Longuet-Higgins MS (1957) The statistical analysis of a random, moving surface, *Philosophical Transactions of the Royal Society of London A*, **249**, 321-387.

[3] Nayak PR (1971) Random process model of rough surfaces, *Journal of Lubrication Technology*, **93**, 398-407.

[4] Whitehouse DJ, Phillips MJ (1978) Discrete properties of random surfaces, *Transactions of*

*the Royal Society of London A*, **290**, 267-298.

[5] Whitehouse DJ, Phillips MJ (1982) Two-dimensional discrete properties of random surfaces, *Transactions of the Royal Society of London A*, **305**, 441-468.

[6] Greenwood JA (1984) A unified theory of rough surfaces, *Proc. Roy. Soc. London A*, **393**, 133-157.

[7] Greenwood JA, Williamson JPB (1966) Contact of nominally flat surfaces, *Proc. Roy. Soc. London A*, **296**, 300-319.

[8] Bush AW, Gibson RD, Thomas TR (1975) The Elastic Contact of a Rough Surface, *Wear*, **35**, 87-111.

[9] Greenwood JA (2006) A simplified elliptic model of rough surface contact, *Wear*, **261**, 191-200.

[10] Majumdar A, Bhushan B (1990) Role of fractal geometry in roughness characterization and contact mechanics of surfaces, *Journal of Tribology*, **112**, 205-216.

[11] Borri-Brunetto M, Carpinteri A, Chiaia B (1999) Scaling phenomena due to fractal contact in concrete and rock fractures, *Int. J. Fracture*, **95**, 221-238.

[12] Persson BNJ, Bucher F, Chiaia B (2002) Elastic contact between randomly rough surfaces: comparison of theory with numerical results, *Phys. Rev. B*, **65**, 184106, 1-7.

[13] Zavarise G, Borri-Brunetto M, Paggi M (2007) On the resolution dependence of micromechanical contact models, *Wear*, **262**, 42-54.

[14] Ausloos M, Berman DH (1985) A multivariate Weierstrass–Mandelbrot function, *Proc. Roy. Soc. London A*, **400**, 331-350.

[15] Ciavarella M, Demelio G, Barber JR, Jang YH (1994) Linear elastic contact of the Weierstrass profile, *Proc. Roy. Soc. Lond. A*, **456**, 387-405.

[16] Wu J-J (2000) Characterization of fractal surfaces, *Wear*, **239**, 36-47.

[17] Paggi M, Ciavarella M (2010) The coefficient of proportionality κ between real contact area and load, with new asperity models, *Wear*, **268**, 1020-1029.

[18] Paggi M, Barber JR (2011) Contact conductance of rough surfaces composed of modified RMD patches, *International Journal of Heat and Mass Transfer*, **54**, 4664-4672.

[19] Pohrt R, Popov VL (2012) Normal contact stiffness of elastic solids with fractal rough surfaces. *Phys. Rev. Lett.*, **108**, 104301.

[20] Paggi M, Pohrt R, Popov VL (2014) Partial-slip frictional response of rough surfaces, *Sci. Rep.*, **4**, 5178.

[21] Zavarise G, Borri-brunetto M, Paggi M (2004) On the reliability of microscopical contact models, *Wear*, **257**, 229-245.

[22] Serge M, Gorb SN (2001) *Biological micro- and nano- tribology*, Nature Solutions Springer.

[23] Varenberg M, Pugno NM, Gorb SN (2010), Spatulate structures in biological fibrillar adhesion, *Soft Matter*, **6**, 3269-3272.

[24] Gao H, Yao H (2004) Shape insensitive optimal adhesion of nanoscale fibrillar structures, *PNAS,* **101**, 7851-7856.

[25] Nosonovsky M, Bhushan B (2007) Hierarchical roughness optimization for biomimetic superhydrophobic surfaces, *Ultramicroscopy*, **107**, 969-979.

[26] Zang J, Ryu S, Pugno N, Wang Q, Tu Q, Buehler MJ, Zhao X (2013) Multifunctionality and

control of the crumpling and unfolding of large-area graphene, *Nature Materials*, **12**, 321-325.

[27] Nosonovsky M, Bhushan B (2008) Capillary effect and instabilities in nanocontacts, *Ultramicroscopy*, **108**, 1181-1185.

[28] Drinkwater BV, Dwyer-Joyce RS, Cawley P (1996) A study of the interaction between ultrasound and a partially contacting solid-solid interface, *Proc. Roy. Soc. London A*, **452**, 2613-2628.

[29] Es F, Demircioglu O, Gunoven M, Kulakci M, Unalan HE, Turan R (2013) Performance of nanowire decorated mono- and multi-crystalline Si solar cells, *Physica E*, **51**:71-74.

[30] Peitgen H, Saupe D (1988) *The Science of Fractal Images* (Berlin: Springer-Verlag).

[31] Thomas TR, Rosén BG (2000) Determination of the optimum sampling interval for rough contact mechanics, *Tribol. Int.*, **33**, 601-610.

[32] Lundberg A J, Wolff L B and Socolinsky D A (2001) New perspectives on geometric reflection theory from rough surfaces, *Proc. of the Eighth IEEE Int. Conf. on Computer Vision (Vancouver, BC, 07-14 Jul 2001)* vol 1 pp 225-32.

[33] Van Ginneken B, Stavridi M, Koenderink JJ (1998) Diffuse and specular reflectance from rough surfaces, *Applied Optics*, **37**, 130-139.

[34] Baker-Finch SC, McIntosh KR (2011) Reflection of normally incident light from silicon solar cells with pyramidal texture, *Progress in Photovoltaics: Research and Applications*, **19**, 406-416.

[35] Acker J, Koschwitz T, Meinel B, Heinemann R, Blocks C (2013) HF/$HNO_3$ etching of the saw damage, *Energy Procedia*, **38**, 223-233.

[36] Koschwitz T, Meinel B, Acker J (2013) Application of confocal microscopy to evaluate the morphology of acidic etched mc-silicon, *Energy Procedia*, **38**, 234-242.

[37] Komatsu Y, Harata D, Schuring EW, Vlooswijk AHG, Katori S, Fujita S, Venema PR, Cesar I (2013) Calibration of electrochemical capacitance-voltage method on pyramid texture surface using scanning electron microscopy, *Energy Procedia*, **38**, 94-100.

[38] Yoshimitsu Z, Nakajima A, Watanabe T, Hashimoto K (2002) Effects of surface structure on the hydrophobicity and sliding behavior of water droplets, *Langmuir*, **18**, 5818-5822.

[39] Bico J, Thiele U, Quéré D (2002) Wetting of textured surfaces, *Colloids and Surfaces A*, **206**, 41-46.

[40] Pugno N M (2007) Mimicking lotus leaves for designing superhydrophobic/hydrophilic and super-attractive/repulsive nanostructured hierarchical surfaces *The Nanomechanics in Italy* Research Signpost, India, pp 1–9.

[41] Kubiak KJ, Wilson MCT, Mathia TG, Carval Ph (2011) Wettability versus roughness of engineering surfaces, *Wear*, **271**, 523-528.

[42] Quéré D (2002) Rough ideas on wetting, *Physica A*, **313**, 32-46.

[43] Duparré A, Ferre-Borrull J, Gliech S, Notni G, Steinert J, Bennett JM (2002) Surface characterization techniques for determining the root-mean-square roughness and power spectral densities of optical components, *Applied Optics*, **41**, 154-171.